\documentclass[12pt]{article}
\usepackage{amsmath}     
\usepackage{citesort}     
\usepackage{amssymb}    
\usepackage{epsfig}  
\def \BE {\begin{equation}}
\def \EE {\end{equation}}
\def \BEA {\begin{eqnarray}}
\def \EEA {\end{eqnarray}}

\begin{document}

\title{Sandpile behavior in discrete water-wave turbulence.}
\author{Sergey Nazarenko
\\ \ \\
{Mathematics Institute,  University of Warwick, Coventry CV4 7AL, UK}}
\maketitle

{\bf Abstract:}
I construct a sandpile model for evolution of the energy spectrum
of the water  waves in finite basins. This model takes
into account loss of resonant wave interactions in discrete 
Fourier space and restoration of these interactions at larger nonlinearity levels.
For weak forcing, the waveaction spectrum takes a critical $\omega^{-10}$ shape where the
nonlinear resonance broadening overcomes the effect of the Fourier grid spacing.
The energy  cascade in this case takes form of rare weak avalanches on the
critical slope background. For larger forcing, this regime is replaced
by a continuous cascade and Zakharov-Filonenko $\omega^{-8}$ waveaction spectrum.
For intermediate forcing levels, both scalings will be relevant, $\omega^{-10}$
at small and $\omega^{-8}$ at large frequencies, with a transitional region
in between characterised by strong avalanches.


\section{Introduction}

Importance of the basin finiteness for the statistical evolution of the
free water surface was recently argued in \cite{naoto,lnp} and \cite{meso}.
Later in the present paper, we will derive an estimate (\ref{a-crit}) according to which 
water surface waves of steepness $\alpha$ 
are sensitive to the basin size $L$ if their wavelength
$\lambda$ is of the order or greater than $ L \alpha^4$.
This means that waves with steepness  $\alpha \sim 0.1$ and 1 meter wavelength
will ``feel'' the boundaries for  for lakes or  gulfs up to
10 km wide.  

Recent direct numerical simulations of the  free water surface in finite basin in presence 
of gravity revealed a bursty character of the nonlinear energy cascade from small to large wavenumbers \cite{lnp}.
Namely, cascade strengths were measured as functions of time at two different wavenumbers within the inertial
range. Intermittent bursts  were observed on these graphs, initially arising at the lower wavenumber 
and then propagating to the higher wavenumber.
This behavior reminds sandpile avalanches moving from low to high wavenumbers and their
mechanism can be understood as a cycle: 
\begin{itemize}
\item A
cascade arrest due to the wavenumber discreteness leads to accumulation of
energy near the forcing scale. 
\item This leads to widening of the
nonlinear resonance. 
\item
Sufficient resonance widening triggers  the
cascade thereby
draining the turbulence levels and returning the system to the
beginning of the cycle. 
\end{itemize}
Note that presence of forcing is essential for this scenario and it should
not be expected in freely decaying fields, e.g. like in \cite{meso}.
Below, we present and study a simple model for such a sandpile-like evolution
of the water wave spectrum.

\section{Differential approximations for waves on an infinite surface.}
Evolution of random weakly nonlinear gravity surface waves in
basins of infinite size and depth is well described by the Hasselmann kinetic equation
\cite{hass_ke}. However, for many purposes one can use a simpler 
differential equation model  which preserves many properties of the
Hasselmann equation \cite{irosh,hass,zp}:
\begin{equation}
\dot n = {C_1 \over g^{3/2} \omega^3} {\partial^2 \over \partial \omega^2}
 n^4 \omega^{26} {\partial^2 \over \partial \omega^2} {1 \over n},
\label{fourth}
\end{equation}
where $C_1$ is a dimensionless constant.
This equation preserves energy
\begin{equation}
E = { 2 \pi  \over g^2} \int \omega^4 n d\omega
\end{equation}
and the waveaction
\begin{equation}
N = {2 \pi   \over g^2} \int \omega^3 n d\omega.
\end{equation}
Even simpler differential approximation was suggested in \cite{zp},
\begin{equation}
\dot n = {C_2 \over g^{3/2} \omega^3} {\partial^2 \over \partial \omega^2} ( n^3 \omega^{24}),
\label{second_zakh}
\end{equation}
where $C_2$ is a dimensionless constant.
This equation conserved both energy and the waveaction, but it does
not have thermodynamic 
solutions corresponding to equipartition of these quantities.

If we insist that having thermodynamics equilibria is important, but
ignore conservation of the waveaction (which can be done at scales smaller,
but not larger, than the
forcing scale) then we can use another
2nd order differential equation,
\begin{equation}
\dot n = {C \over g^{3/2} \omega^4} {\partial \over \partial \omega} \left(n^2 \omega^{24}
{\partial \over \partial \omega} (\omega n) \right),
\label{second}
\end{equation}
where $C$ is a dimensionless constant.
Similar approach in Navier-Stokes turbulence is called Leith model
\cite{Leith,cn}. Like in Leith model \cite{cn}, we can now find a ``warm cascade''
solution, i.e. the general stationary solution of (\ref{second}) which
contains both finite flux and finite temperature components,
\begin{equation}
n = {1 \over  \omega} \left(  {g^{3/2} P \over 7 C}  \omega^{-21} +T^3 \right)^{1/3}
\label{warm}
\end{equation}
where $P$ and $T$ are (dimensional) constants measuring the energy flux and
the temperature respectively. For  $T=0$ 
we recover the pure cascade Zakharov-Filonenko state,
\begin{equation}
n =  \left(  {P / 7C }   \right)^{1/3}   g^{1/2} \omega^{-8}, 
\label{zf}
\end{equation}
 and for $P=0$
we get the pure thermodynamic distribution,
\begin{equation}
n = { T \over  \omega}.
\label{term}
\end{equation}

\section{Finite-basin effects.}

Let us now consider waves in a square basin with sides
of length $2 \pi$, so that the wavenumbers take
values on a discrete lattice, ${\bf k} \in {\cal Z}^2$.
The main effect of the finite basin size is in loss
of wavenumber resonances due to the wavenumber discreteness
\cite{naoto,lnp,meso}.
Indeed, let us consider  the 4-wave
resonance conditions,
\begin{eqnarray}
{\bf k}_1 + {\bf k}_2 &=& {\bf k}_3 + {\bf k}_4, \\
\omega({\bf k}_1) + \omega({\bf k}_2)
 &=& \omega({\bf k}_3) + \omega({\bf k}_4).
\end{eqnarray}
Two different classes of such solutions were found in \cite{lnp}:
collinear quartets (all four wavevectors are parallel to each other)
and ``tridents'' (first two wavevectors are anti-parallel 
 and the other two are mirror symmetric with respect to the direction of the
first two). Parametrisation of the collinear quartets and quintets was done
in \cite{dlz}. Note that the collinear quartets are physically unimportant
because of the zero nonlinear coefficient for such wavevectors, but
the next order (5-wave) is nontrivial and yields an interesting
kinetic equation \cite{dlz}. The second class, tridents, can be paramertised
as follows \cite{lnp},
$$\mathbf{k}=(a,0),\;\;  \mathbf{k}_1=(-b,0),\;\;
\mathbf{k}_2=(c,d),\;\; \mathbf{k}_3=(c,-d) $$
with
$$ a = (l^2+m^2+lm)^2, \;\;
b = (l^2+m^2-lm)^2,  \;\;
c= 2lm(l^2+m^2), \;\;
d= l^4-m^4,
$$
where $l$ and $m$ are integers.
New solutions can be obtained by further rescaling and rotating
these tridents by rational angles.

Further, more resonances appear due to the nonlinear resonance 
broadening even when this broadening is significantly less
than the wavenumber grid spacing \cite{lnp}. 
However, the total number of both exact and approximate resonances
remain significantly depleted with respect to the continuous case
and, therefore, they are inefficient for supporting the 
turbulent cascade unless the nonlinear resonance broadening
becomes of order of the $k$-grid spacing.
This allows us to formulate a simplified
model for wave turbulence in finite basins as explained in the next section.

\section{Wave turbulence in finite basins.}

First, we need to evaluate the 4-wave resonance broadening which is of order of
the characteristic nonlinear time $\tau_{NL}$. Estimate for $\tau_{NL}$  will be the
same if one finds it using (\ref{fourth}), (\ref{second_zakh}),
(\ref{second}) or the original Hasselmann's kinetic equation
\cite{hass_ke}. It can also be found from a simple dimensional argument
and the result is
\begin{equation}
 \tau_{NL} \sim g^{10} \omega^{-19} n^{-2}.
\label{tau}
\end{equation}
 This corresponds to the resonance broadening in the $k$-space given by
\begin{equation}
 \kappa_{NL} ={1 \over {\partial \omega \over \partial k} \tau_{NL}} \sim g^{-11} \omega^{20} n^{2}.
\label{kappa}
\end{equation}
In our model, we will postulate that the wave spectrum will not evolve at $\omega$
if the resonance broadening $\kappa_{NL}$ is less that the $k$-grid spacing
$\kappa$,
and it will evolve as in the continuous case for   $\kappa_{NL} > \kappa$.
Such a ``frozen turbulence'' state was first observed in numerical
simulations of the capillary waves \cite{frozen} and it was later
discussed in \cite{CNP,lnp}. One has to be careful, however, not to interpret
literally the absence of the spectrum evolution at small amplitudes because
a small number of exact resonances do survive for the gravity waver waves 
(see the previous section) 
and further  resonances may re-appear at resonance broadening
which is much less than the $k$-grid spacing \cite{lnp}. However, the number
of such resonant modes is too small to evolve the spectrum efficiently and
in our simple model we just
put the Heaviside step function $H(\kappa_{NL} - \kappa)$ as a pre-factor to the 
equation (\ref{second}) in order to get a model for the spectrum
evolution in discrete  $k$-space,
\begin{equation}
\dot n = {H(\kappa_{NL} - \kappa) \over g^{3/2} \omega^4}
 {\partial \over \partial \omega} \left(n^2 \omega^{24}
{\partial \over \partial \omega} (\omega n) \right) + \gamma(\omega) n.
\label{discrete}
\end{equation}
Here, we have added a function $\gamma(\omega)$ which models forcing
at low $\omega$'s (e.g. by wind) and dissipation at high $\omega$'s
(by wavebreaking) and which, in principle, can be a
function of $n$. We will assume that in between of the
forcing and dissipation scales there exists an inertial range where 
$\gamma \approx 0$.
 From now on, we ignore the
dimensionless order-one pre-factor $C$ since our resonance broadening is
given by an order-of-magnitude estimate.

\section{Behavior predicted by the model.}
Equation (\ref{discrete}) with  $\kappa_{NL}$ given by (\ref{kappa}) will
be our master model for the water-wave turbulence spectrum in a finite basin.
Let us qualitatively consider the consequences of this model.
Let us assume that initially there is no waves in the basin and let us
start forcing the system at low frequencies. Then, there will be no transfer
over scales and the spectrum will grow with the growth rate $\gamma$ until
it reaches the critical value where  $\kappa_{NL} = \kappa$. After that
the nonlinear transfer will get activated and the energy will spill into the
adjacent range of frequencies. If the forcing is so weak that its
characteristic $\tau_F = 1/\hbox{max} \{\gamma(\omega) \}$ is much longer than
the characteristic nonlinear time $\tau_{NL}$ then the level of turbulence
will never greatly exceed its critical value 
and the critical spectrum with   $\kappa_{NL} \approx \kappa$ will gradually
occupy the entire inertial range. Condition $\kappa_{NL} \approx \kappa$ 
gives for the critical spectrum
\begin{equation}
 n_c \sim g^{11/2}  \kappa^{1/2}  \omega^{-10}.
\label{crit}
\end{equation}
It is useful to re-write this relation in terms of the
water surface angle $\alpha$ characterising the wave steepness,
\begin{equation}
 \alpha_c \sim (\lambda /L)^{1/4},
\label{a-crit}
\end{equation}
where $L$ is the box size and $\lambda$ is the wavelength.
One can interpret this relation as an expression for the minimal
 steepness for which the finite box effect can be ignored.
For example, for a box containing $10^4$ wavelengths the finite box
effects can be ignored only for $\alpha > 0.1$.

 If the forcing is stochastic then the
subsequent evolution will consist of small avalanches going down the 
critical slope with time intervals $\Delta t$ greater than the time $\sim
\tau_{NL}$
 needed (according to (\ref{discrete}))
 for the avalanche to travel from the forcing to the dissipation scale.
Note that this is a classical condition for sandpile models.
 To be specific, let us consider a type of forcing such that
after each interval $ \Delta t$ we add an increment $(\Delta a) e^\phi$
where phase $\phi$ is random and uniform in $(0, 2 \pi]$
and $a$ is a small positive value, $\Delta a \ll \sqrt n_c$.
If at some moment of time $n=n_c$ at the forcing scales $k\in (k_F, k_F+\Delta
k)$
then with probability $1/2$ the spectrum will get greater
than critical in the forcing range after time interval $\Delta t$ and
approximately $n= n_c + \Delta a  \sqrt n_c$.
For $\Delta t \gg \tau_{NL} $, such disturbance will have 
enough time to travel/diffuse away before the next spectrum
disturbance might appear at $k_F$ after another $\Delta t$ interval.
Thus, the evolution of each super-critical disturbance can
be treated separately. Because each of such disturbances
is small, one can use the linearised version of
the evolution equation 
(\ref{discrete}),
\begin{equation}
\dot n = \kappa  g^{19/2} \omega^{-4}
 {\partial \over \partial \omega} \left( \omega^{4}
{\partial \over \partial \omega} (\omega n) \right),
\label{linear}
\end{equation}
with initial condition
\begin{equation}
n|_{t=0} =\Delta
 a  \sqrt { n_c(\omega_F)} \;\;\;\; \hbox{for} \;\; 
\omega \in \omega_F + \Delta \omega, \;\;
 \hbox{and} \;\; n=0 \;\; \hbox{otherwise}.
\label{ic}
\end{equation}
Equation (\ref{linear})  can be re-written as 
\begin{equation}
\dot \epsilon = \kappa  g^{19/2} \left(4{\partial \epsilon \over \partial \omega} 
+\omega{\partial^2  \epsilon \over \partial \omega^2} \right),
\label{linear1}
\end{equation}
where $\epsilon = \omega n$ is the spectral energy density.
According to this equation, the disturbance generated by forcing will
propagate toward high $k$ with speed $4\kappa  g^{19/2}$ while getting
diffused at an increased rate (due to moving to higher $\omega$'s).
The stationary solution of (\ref{linear1}) decays as $\epsilon \sim
\omega^{-3}$ or  $n \sim
\omega^{-4}$, i.e. significantly slower than $n_c \sim \omega^{-10}$.
 Therefore, for a long enough inertial range the linear approximation
will fail at some $\omega = \omega^*$ and the critical
slope $n \sim \omega^{-10}$ will be replaces by the Zakharov-Filonenko slope
 $n \sim \omega^{-8}$ for $\omega > \omega^*$.
The transitional range with $\omega \sim \omega^*$ will be characterised by
strong avalanches.
 
For stronger forcing, transition to the  Zakharov-Filonenko spectrum
occurs at lower frequencies or even right at the forcing scale
if   $\omega_F > \omega^*$ (i.e. when $\alpha$ at the forcing scale
is steeper than $\alpha_c$ at this scale). 
In numerical simulations, efforts are typically made to
overcome ``frozen turbulence'' and generate the cascade.
At the present level of resolution (up to $512^2$ modes) this
goal can be achieved with only partial success because,
according to estimate (\ref{a-crit}), the condition that turbulence is
not frozen can be only be marginally reconciled with the condition
for the Wave Turbulence theory to work, $\alpha < 1$.
Thus, in all existing simulations (e.g. \cite{onor,naoto,naoto1,PRZ,lnp,meso}) 
turbulence, although not frozen, was
still quite sensitive to the finite box effects. This state was named
mesoscopic turbulence in \cite{meso}.   In presence of forcing, it shows up via strong 
cascade avalanches coexisting with  Zakharov-Filonenko
state occupying about a decade long wavenumber interval.  

\section{Discussion}

In this paper, we presented an evolution model for the
spectrum of gravity water waves in finite basins. It has the following 
features:
\begin{itemize}
\item The model is give by a nonlinear second order
equation in Fourier space.
\item The model has the cascade Zakharov-Filonenko spectrum
and the thermodynamic spectrum among its solutions.
It also has a general stationary solution where both
the flux and the temperature effects are present.
\item The model takes into account the $k$-space
discreteness by switching off the nonlinear 
evolution when the spectrum falls below a critical
value. The critical spectrum is determined by the condition
that the nonlinear resonance widening is equal to the 
$k$-grid spacing.
\end{itemize}

Based on this model, we established that for very weak
forcing the spectrum takes the critical slope $n \sim
\omega^{-10}$ with occasional weak ``avalanches'' running
down this slope. For larger forcing the system does not
feel discreteness and the spectrum takes the 
Zakharov-Filonenko form, $n \sim \omega^{-8}$.
For intermediate levels of forcing, the spectrum may have
the $-10$ exponent at low frequencies and the $-8$
 at large frequencies within the inertial range. Such intermediate
case is characterised by strong avalanches down the mean spectral slope,
a feature observed in recent numerical simulations of the free water
surface \cite{lnp}.

It is interesting that steeper than Kolomogorov slopes were also previously
obtained for waves with narrow-band forcing \cite{zlf}. The narrow band forcing
also leads to quasi-discrete character of the mode excitations. 
To conclude, it is worth mentioning that there are plenty of other
examples of dispersive waves whose resonant interaction may be 
affected by the finite size box and where one could expect similar
avalanche-like behaviour. Interestingly, irrespective to discreteness,
the sandpile analogy have also been previously invoked in the wave turbulence context
to illustrate sudden readjustments necessary to balance the turbulence sources and sinks
\cite{newell}.

\end{document}